\documentclass[twocolumn,superscriptaddress,showpacs,prl]{revtex4}

\newcommand{\chalmers}{Department of Applied Physics,
Chalmers University of Technology,
SE-412\;96 G\"{o}teborg, Sweden}
\newcommand{\rutgers}{Department of Physics and Astronomy, Rutgers University, 
Piscataway, N.\ J.\ 08854-8019}

\usepackage{graphicx}
\usepackage{amsmath}
\usepackage{dcolumn}
\usepackage{bm}

\renewcommand{\vec}[1]{\mathbf{#1}}
\begin{document}

\title{Towards a working density-functional theory for polymers:\\
First-principles determination of the polyethylene crystal structure}

\author{Jesper Kleis}\affiliation{\chalmers}

\author{Bengt I. Lundqvist}\affiliation{\chalmers}

\author{David C. Langreth}\affiliation{\rutgers}

\author{Elsebeth Schr{\"o}der}\affiliation{\chalmers}

\date{\today}
\begin{abstract}

Equilibrium polyethylene crystal structure, cohesive
energy, and elastic constants are calculated by density-functional
theory applied with a recently proposed density functional
(vdW-DF) for general geometries [Phys.\ Rev.\ Lett.\ \textbf{92},
246401 (2004)] and with a
pseudopotential-planewave scheme. The vdW-DF with its account 
for the long-ranged van der Waals interactions gives
not only a stabilized crystal structure but also values of the
calculated lattice parameters and elastic constants in quite good
agreement with experimental data, giving promise for successful
application to a wider range of polymers.
\end{abstract}


\maketitle

Understanding of crystalline solids is greatly enhanced by 
the periodicity of the atomic structure, which allows very 
detailed comparisons between theory and experiment. 
Macromolecular materials, such as polymers and other complex 
fluids, are less well understood, due to the challenging complex 
natures of their structures. Features of the polymer on the atomic 
and mesoscopic length scales are inter-dependent. For instance, 
atomic bonds are on the \AA ngstr{\"o}m scale,  while diffusion 
processes involve whole chains extending some 100 {\AA}. In 
order to cover the full set of length scales by theory we need 
atomic-scale input for mesoscopic-scale force-scheme simulations. 
Input forces may be determined routinely from first-principles 
density-functional theory (DFT) calculations, the more accurate a
functional the better, for a reasonably-sized 
unit cell. Among polymer crystals, with their typically very 
complex structure, the polyethylene (PE) crystal has a relatively
simple structure~\cite{Dorset}. A 
first-principles 
implementation of DFT~\cite{gg}, vdW-DF, with a general 
account for van der Waals (vdW) forces, is here shown to give 
results for crystal structure, cohesive energy, and elastic constants 
of the PE  crystal in a very promising agreement with previous 
low-temperature measurements.

Earlier attempts to study the equilibrium properties of polymer 
systems have relied on semiempirical interatomic potentials, 
including force-field methods%
~\cite{PEforcefieldAnalysis,PolymerForceFieldSorensen} and 
modified DFT calculations~\cite{DFTnotLDASerra}. The fitting of 
internuclear potentials in order to reproduce experimental equilibrium 
properties gives serious restrictions on the predictive power of these 
methods for nonequilibrium situations and for molecules, for which 
experimental data are scarce or lacking. Further, these semiempirical 
methods lack information to make a systematic improvement of the 
interatomic potentials possible. A physically motivated first-principle 
description of sparse polymer crystals is highly desirable. When 
applied at perfect conditions, it might even render the possibility of 
developing consistent and transferable interatomic potentials for both 
force schemes and hybrid methods.

The ongoing DFT success story for ground-state electron structure 
calculations of dense materials systems is driven by the relatively 
low computational cost together with the ability to describe  very 
diverse systems. 
However, the widely used local and semilocal 
implementations do not include long-range nonlocal correlations
that are essential for a proper description of the intermolecular 
vdW interactions. These interactions are crucial for the stability 
of systems with regions of low electron density, often encountered 
in biological and nanotechnological applications.
The recently developed vdW-DF 
functional~\cite{gg} accounts for the vdW interaction in a 
seamless way both at the equilibrium binding separation and at 
asymptotically large separations. It applies to general geometries 
and accounts for the nonlocal correlation 
via a functional that 
takes only the electron density as input. It has proved very 
promising in describing equilibrium separations and binding 
energies for a range of systems, including dimers of benzene 
rings~\cite{gg,Aron2}, graphene sheets~\cite{bonog}, 
polycyclic aromatic hydrocarbons (PAH's)~\cite{SvetlaPAH2}, 
monosubstituted benzene molecules~\cite{timo}, and combinations 
thereof, such as PAH's and phenol adsorbed on graphite~\cite{bonog,phenol}.

\begin{figure}
\centering
\includegraphics[width=0.5\columnwidth]{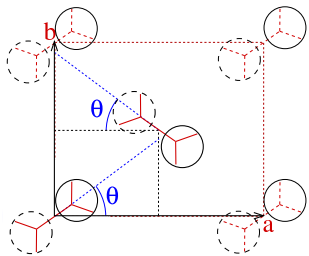}
\caption{\label{fig:SettingAngle}
Schematics of the PE crystal structure in its base-centered 
orthorhombic unit cell with the polymers aligned in parallel. 
Lattice parameters $a$ and $b$ are indicated, while the 
parameter $c$, which is the polymer repetition length, is 
perpendicular to the plane shown. Solid (broken) circles 
represent CH$_2$-units in the plane (a distance $c/2$ out 
of the plane). The angular orientation of the polymers is 
determined by the setting angle $\theta$, defined as the angle 
between the $a(>b)$ axis and the intra-polymer carbon plane.
}
\end{figure}

Simple polymer crystals constructed from long parallel polymer 
chains~\cite{Dorset}, such as linear PE or polypropylene, 
represent an important class of sparse materials that are 
stabilized by the vdW interaction. At low temperatures, the 
PE crystal stabilizes in a simple base-centered orthorhombic 
crystal structure (Pna$2_1$), Fig.~\ref{fig:SettingAngle}, 
as established by X-ray and neutron scattering experiments%
~\cite{PEXrayBunn,ActaCryst2,ExpPolyEthylenexrayc,NeutronPolyethylene}. 
Standard theory, first-principles calculations with the semilocal 
generalized gradient approximation (GGA) of DFT, predicts this 
structure to be unstable \cite{DFTnotLDASerra,DFTPolyethyleneLett}, 
calling for incorporation of nonlocal interactions. Recent vdW-DF 
calculations on parallel, well separated PE 
molecules~\cite{kleisEMRS,kleisJCP} show the vdW interaction in these 
systems to be non-negligible. To account for the nonlocal correlations 
at all separations, in particular those relevant in the PE crystal, we 
here apply the general-geometry vdW-DF~\cite{gg}.

The structure of the isolated polymer has earlier been 
calculated in the GGA~\cite{kleisJCP}, and this structure 
is used as input for the crystal calculations \cite{footnote1}. 
The internal 
geometrical structure of the isolated polymer chain thus fixes 
the repetition length $c=2.57$~{\AA} of the PE crystal unit cell.

Our vdW-DF calculations give binding-energy values at varying 
crystal-parameter settings, which are presented as contour maps, 
and from which equilibrium values for lattice parameters, 
cohesive energies, and elastic coefficients are extracted and 
compared with experimental low-temperature data. Polymers 
might have several conformations with similar energies. 
Likewise, in a polymer crystal structure several local energy 
minima might exist. In particular, whereas the structure shown 
in Fig.~\ref{fig:SettingAngle} with $a>b$ is experimentally 
observed, the structure that approximately corresponds to 
pulling one of the polymers half a unit length ($c/2$) out of 
the plane of the paper leads to essentially the same polymer 
chain packing and in turn to a similar configurational energy. 
In fact, the structure with a polymer pulled outward can 
also be described by the schematics in Fig.~\ref{fig:SettingAngle}, 
when the requirement $a>b$ is relaxed. Our first-principles 
method enables us to directly assess the cohesive-energy 
contours in both ($a$, $b$, $\theta$) regions.

\begin{figure}
\centering
\includegraphics[width=0.8\columnwidth]{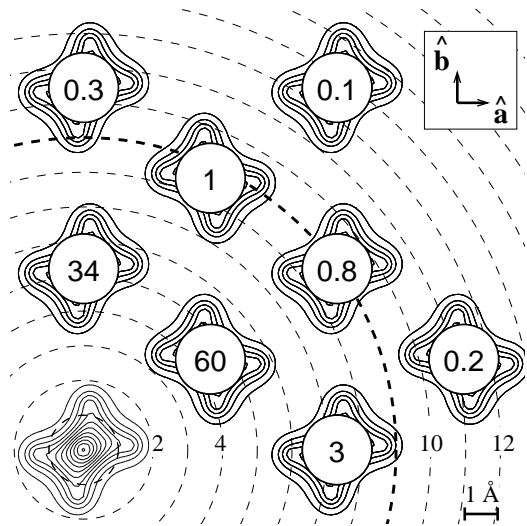}
\caption{\label{fig:QuartCryst} 
The nonlocal correlation energy $E_\mathrm{c}^\mathrm{nl}$ 
for pairs of PE. For the practical 
evaluation of $E_\mathrm{c}^\mathrm{nl}$ in the crystal the range of integration 
in Eq.\ (\ref{eq:Ecnl}) is important. Simple pair interaction 
calculations give a fast estimate of the range needed, although 
the full calculation includes all density within the integration 
range. 
Shown here are the relative sizes (meV per polymer segment) 
of the polymers' 
contributions to  
$E_\mathrm{c}^\mathrm{nl}$ (as pairs with the polymer at the circle centers), 
evaluated at the separations ({\AA}) and relative angles shown. 
The contours show length-averaged 
electron densities.
Polymers that are closer than 9~{\AA} (thick dashed circle) to the 
center polymer contribute with an energy of 321 meV. This is 
99.4\% of the total nonlocal correlation energy  $E_\mathrm{c}^\mathrm{nl}$ 
of all the  polymers shown (323 meV).
}
\end{figure}

The vdW-DF functional is derived and described in 
Ref.~\onlinecite{gg}. It divides the correlation-energy functional
into local and nonlocal parts,
\begin{equation}
E_\mathrm{c}
\approx E_\mathrm{c}^{\mathrm{LDA}}+E_\mathrm{c}^{\mathrm{nl}}\, ,
\end{equation}
where the first term is approximated by the local-density
approximation (LDA) and the second term vanishes for a
uniform system. The dispersive interactions give the second term
a substantial nonlocality, which allows a simpler account
of the polarization properties from which it originates.
It is determined from the inverse dielectric function
$\epsilon^{-1}$, which is assumed to be dominated by a single pole,
with a strength fixed by the f-sum rule. The functional form
obtained approximately is
\begin{equation}
E_\mathrm{c}^\mathrm{nl}[n] = \frac{1}{2}\int d\vec r \, \int d \vec r'
n(\vec r) \phi(\vec r, \vec r') n(\vec r') \, .\label{eq:Ecnl}
\end{equation}
The nonlocal kernel $\phi$ can be tabulated \cite{gg} in
terms of a dimensionless distance
$D=(q_0 + q'_0)|\vec r - \vec r'|/2$ and an asymmetry parameter
$\delta = (q_0 - q'_0)/(q_0 + q'_0)$, where $q_0$ is a local
parameter that depends on the electron density and its
gradient at position $\vec r$. The quantity $q_0$ is
related to the pole position in $\epsilon^{-1}$, which is
determined by the requirement that this $\epsilon$
should give the same appropriate semilocal exchange-correlation
energy density component as that of an electron gas in a
much better approximation.

The nonlocal-energy integral (\ref{eq:Ecnl}) has the 
electronic density $n(\vec r)$ as input. We use a 
self-consistently determined GGA density with the revPBE 
exchange flavor~\cite{revPBE,gg,ijqc}, calculated for the 
full crystal unit cell using a plane-wave code~\cite{dacapo} 
with ultrasoft pseudopotentials~\cite{vdBildt}. The 
plane-wave basis set is truncated at 400~eV, and a 
($4\times 4\times 10$) $k$-point
Monkhorst-Pack 
sampling of the Brillouin zone is used for the 
the periodically repeated unit cell.
A spatial sampling separation of 0.13 {\AA} between 
fast-Fourier-transform grid points is used.

In Ref.~\onlinecite{gg} the vdW-DF is 
described for and applied to finite molecules. The generalization 
to extended systems is straightforward. For an adsorbate system 
(a finite molecule plus an extended surface) this is described in 
Ref.~\onlinecite{bonog}. For a bulk system, like the PE crystal, 
$E_\mathrm{c}^\mathrm{nl}[n]$ of the crystal unit cell must 
include the interactions from the surrounding cells. This is 
taken care of by letting the spatial integrals in Eq.~(\ref{eq:Ecnl}) 
(as energy per unit cell) extend over the unit cell for $\vec r$ and 
``everywhere'' for $\vec r'$. In practice, the $\vec r'$ integral is 
carried out over a region in space that is sufficiently large, 
i.e.\ by adding more space we do not change 
$E_\mathrm{c}^\mathrm{nl}$ at the desired level of accuracy. 
We find it sufficient to include a total of 5 unit cells in each of 
the $a$ and $b$ directions (Fig.~\ref{fig:SettingAngle}) and 9 
unit cells in the $c$ direction in the $\vec r'$ integral, 
by which $E_\mathrm{c}^\mathrm{nl}$ is found to deviate less 
than one part per thousand from the 
results obtained by integrating over a substantially larger 
region of space. Figure~\ref{fig:QuartCryst} illustrates this by 
showing the pairwise contributions to the polymer interaction 
$E^{\text{nl}}_{\text{c}}$. The 
figure shows that the most important contributions come from 
the nearest and next-nearest neighbor polymers.

The total energy of the crystal is thus
\begin{equation}
  E_{\text{vdW-DF}} =  E_{\text{GGA}} - E_{\text{GGA,c}} 
+ E_{\text{LDA,c}} + E^{\text{nl}}_{\text{c}}\, ,
\label{eq:vdW-DF}
\end{equation}
with the revPBE flavor of GGA. The cohesive energy is 
obtained as the difference between the calculated total-energy 
values for the actual structure and for a reference structure 
with widely separated PE polymers. The electron-density-grid 
spacing as well as the polymer position relative to the grid are 
kept fixed for the vdW-DF reference calculation to ensure that 
any polymer self-interaction of the crystal calculation and 
most grid-related numerical errors are canceled by the 
reference calculation. The scheme applies also to other 
sparse-matter systems.

\begin{figure}
\centering
\includegraphics[width=0.95\columnwidth]{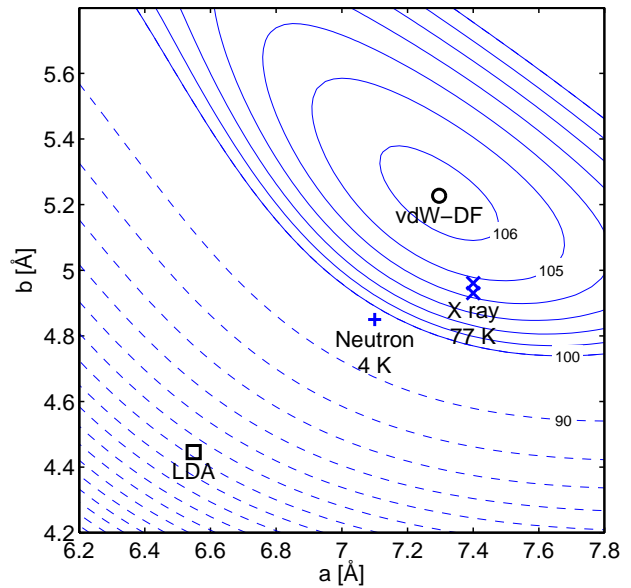}
\caption{\label{fig:CohesiveEnergyLandscape} 
The cohesive-energy contour map calculated with vdW-DF, 
expressed as energy per CH$_2$ group as a function of the 
lattice parameters $a$ and $b$. The plot (but not the 
calculation) is restricted to show the energetics at setting 
angle $\theta=44^\circ$, 
which is the optimal angle found with vdW-DF (Table~\ref{tab:el}). 
The unit-cell height is fixed to $c=2.57$ {\AA}. The full (dashed) 
contour curves are separated by 1 (10) meV/CH$_2$, with values 
given in meV/CH$_2$ at selected contours. The equilibrium 
parameters (summarized in Table~\ref{tab:el}) obtained from 
experiments and from a LDA-DFT study are shown for comparison.}
\end{figure}

\begin{table*}
\begin{center}
\caption{\label{tab:el}
Values of equilibrium lattice parameters $a_0$ and $b_0$, setting 
angle $\theta_0$, elastic constants $C_{11}$, $C_{12}$, and 
$C_{22}$, and cohesive energy $E_\mathrm{coh}$ per CH$_2$ 
group, calculated by the vdW-DF method and compared with 
other theoretical and experimental data.}
\begin{tabular}{lccccccc}
\hline \hline
Method &$C_{11}$ (GPa)&$C_{12}$ (GPa)&$C_{22}$ (GPa)&$E_\mathrm{coh}$ 
(eV/CH$_2$)&$a_0$ (\AA)&$b_0$ (\AA)&$\theta_0$ (degr.)\\
\hline
vdW-DF                 & 11.1 &  6.3 &  8.9 & 0.106$^j$ & 7.30 & 5.22 & 44  \\
vdW-DF                 &  8.8 &  6.7 & 12.8 & 0.105$^j$ & 5.29 & 7.28 & 51  \\
LDA$^a$                & 41.2 & 11.3 & 47.1 & 0.11$^j$  & 6.55 & 4.45 & 44.4\\
LDA+BLYP+6-LJ$^b$      & 14.1 &  7.2 & 11.8 & 0.11$^j$  & 7.40 & 4.90 & 45  \\
Force field (4 K)$^c$  & 14.0 &  7.9 & 13.5 & 0.081     & 7.20 & 4.80 & 41.9\\
Neutrons (77K)$^d$     & 11.5 &      &      &           &      &      &     \\
Experiment$^e$         &      &      &      & 0.080     &      &      &     \\
X rays (77 K)$^f$      &      &      &      &           & 7.42 & 4.96 & 47.7\\ 
X rays (77 K)$^g$      &      &      &      &           & 7.39 & 4.93 & 45  \\
Neutrons (4 K)$^h$     &      &      &      &           & 7.12 & 4.85 & 41  \\
Experiment (213 K)$^i$ &  8.4 &  4.2 &      &           &      &      &     \\
\hline \hline
\multicolumn{8}{l}{%
$^a$Ref.~\onlinecite{PECrystalwithPressure}.
$^b$Ref.~\onlinecite{DFTnotLDASerra}.
$^c$Ref.~\onlinecite{PEforcefieldAnalysis}.
$^d$Ref.~\onlinecite{ExpPolyEthylene}.
$^e$Ref.~\onlinecite{Bill}.
$^f$Ref.~\onlinecite{ActaCryst2}.
$^g$Ref.~\onlinecite{ExpPolyEthylenexrayc}.
$^h$Ref.~\onlinecite{NeutronPolyethylene}.
$^i$Ref.~\onlinecite{ChoyElasticPE}.}\\
\multicolumn{8}{l}{%
$^j$Value not corrected for zero-point motion, estimated to account for approximately 0.01 eV/CH$_2$.} 
\end{tabular}
\end{center}
\end{table*}

The cohesive-energy contours of the PE-crystal, that is 
the loci of equal binding-energy values for configurations 
with varying values of the lattice parameters $a$, $b$ and 
the setting angle $\theta$ (Fig.~\ref{fig:CohesiveEnergyLandscape}), 
are calculated with vdW-DF, and two nearly equally deep minima 
are found at two different setting-angle values (Table~\ref{tab:el}). 
The configuration approximately sketched in 
Fig.~\ref{fig:SettingAngle} is the most stable of the two. 
However, the two configurations are very close in energy, 
differing by only 1 meV/CH$_2$ (1\%), and an accuracy 
sufficiently high to resolve such a stability issue cannot be 
claimed for the vdW-DF method. 

The calculated values of equilibrium lattice parameters for the 
Fig.~\ref{fig:SettingAngle} structure are in close agreement 
with the experimental ones (Table \ref{tab:el}), the vdW-DF 
values $(a_0,b_0) = (7.30\mbox{ {\AA}}, 5.22\mbox{ {\AA}})$ 
differing by at most 8\% from the experimentally observed 
values. The calculated values of the cohesive energy (not 
including zero-point fluctuations) and the elastic coefficients 
(Table~\ref{tab:el}) are also in good accord with experiments.

In comparison with other attempts to calculate the PE-crystal 
equilibrium data (Table~\ref{tab:el}), the vdW-DF results 
show a very promising agreement with experimental data. 
This is, indeed, in great contrast to the fact that the PE-crystal 
is unstable in GGA, according to earlier 
studies~\cite{DFTnotLDASerra,PECrystalwithPressure,DFTPolyethyleneLett} 
and confirmed here, and also to the clear overbinding of the 
LDA~\cite{PECrystalwithPressure}. Calculations with 
ad-hoc corrections for the dispersive interactions~\cite{DampPot} 
with the BLYP functional~\cite{DFTnotLDASerra} and 
with the force-field method~\cite{PEforcefieldAnalysis} give 
results in a fair agreement with experimental data. However, 
the former uses an empirically damped $-R^{-6}$ internuclear 
potential~\cite{DFTnotLDASerra}, and the latter 
has parameters fitted to experimental data at 4 K, such as 
lattice parameters~\cite{PEforcefieldAnalysis}, providing agreement 
with experiment at this temperature by construction. As a 
contrast, the vdW-DF calculations do not take any empirical input.

The vdW-DF functional used here combines the correlation-energy 
functional $E_\mathrm{c}^{\mathrm{nl}}$ (Eq.~(\ref{eq:Ecnl})) 
with an exchange functional taken from the revPBE flavor of GGA. 
Although the latter closely represents the Hartree-Fock exchange 
at separations relevant for this work, it gives a slight 
misrepresentation of the true exchange repulsion~\cite{gg,timo,Aron2}. 
An analysis of the PE-molecule density profile implies that this effect 
should be stronger in the $b$ than in the $a$ direction, which is 
indeed the case for our calculated results. Thus an improved 
exchange description should give an even better agreement with the 
experimental geometry.

In summary, the vdW-DF method successfully predicts structural, 
cohesive, and elasticity data for the important test case of the PE 
crystal. It is gratifying that the inclusion of the fully nonlocal 
vdW-DF correlation functional [Eq.~(\ref{eq:Ecnl})], which has 
no empirical input or fitted parameters, leads to a good agreement 
with experimental data. As similar conclusions are drawn for a 
variety of different carbon-based systems, like graphene, dimers 
of benzene, polyaromatic hydrocarbons, monosubstituted benzene 
molecules, and combinations 
thereof~\cite{bonog,Aron2,timo,SvetlaPAH2,phenol}, the vdW-DF functional 
is certainly very promising for general kinds of geometries and 
molecules, like those in macromolecular materials. 

Support from the Swedish Research Council and the Swedish 
National Graduate School in Materials Science, as well as 
allocation of computer resources at UNICC (Chalmers) and 
SNIC (Swedish National Infrastructure for Computing), is 
gratefully acknowledged. Work by D.C.L.\ was supported in 
part by NSF Grant DMR-0456937.


\begin{thebibliography}{99}

\bibitem{Dorset}
D.L. Dorset,
Rep. Prog. Phys. \textbf{66}, 305 (2003).

\bibitem{gg}
M. Dion et al., 
Phys. Rev. Lett. \textbf{92}, 246401 (2004); \textbf{95}, 109902(E) (2005).

\bibitem{PEforcefieldAnalysis}
N. Karasawa, S. Dasgupta, and W.A. Goddard III,
J. Phys. Chem. \textbf{91}, 2263 (1991).

\bibitem{PolymerForceFieldSorensen}
R. A. Sorensen et al., 
Macromol. \textbf{21}, 200 (1988).

\bibitem{DFTnotLDASerra}
S. Serra et al., 
Chem. Phys. Lett. \textbf{331}, 339 (2000).

\bibitem{Aron2}
A. Puzder, M. Dion, and  D.C. Langreth,
J. Chem. Phys. \textbf{124}, 164105 (2006).

\bibitem{bonog}
S.D. Chakarova-K\"ack et al., 
Phys. Rev. Lett. \textbf{96}, 146107 (2006).

\bibitem{SvetlaPAH2}
S.D. Chakarova-K\"ack, J. Kleis,  and E. Schr\"oder,
\textit{Dimers of polycyclic aromatic hydrocarbons in
density functional theory}, Applied Physics Report 2005-16.

\bibitem{timo}
T. Thonhauser, A. Puzder, and D.C. Langreth,
J. Chem. Phys. \textbf{124}, 164106 (2006).

\bibitem{phenol}
S.D. Chakarova-K\"ack et al.,
Phys. Rev. B. \textbf{74}, 155402 (2006).

\bibitem{PEXrayBunn} 
C.W. Bunn, Trans. Faraday Soc. \textbf{35},
482 (1939).

\bibitem{ActaCryst2}
P.W. Teare, Acta Cryst. \textbf{12}, 294 (1959).

\bibitem{ExpPolyEthylenexrayc}
S. Kavesh and J.M. Schultz, J. Polym. Sci. A-2 \textbf{8},
243 (1970).

\bibitem{NeutronPolyethylene}
G. Avitabile et al.,
J. Polym. Sci., Polym. Lett. Ed. \textbf{13}, 351 (1975).

\bibitem{DFTPolyethyleneLett}
B. Montanari and R.~O. Jones, Chem. Phys. Lett. \textbf{272},
347 (1997).

\bibitem{kleisEMRS}
J. Kleis, P. Hyldgaard, and E. Schr\"oder, 
Comp. Mat. Sci. \textbf{33}, 192 (2005).

\bibitem{kleisJCP}
J. Kleis and E. Schr\"oder, J. Chem. Phys. \textbf{122},
164902 (2005).

\bibitem{footnote1}
We find that for separations typical in the PE crystal, the influence 
on the internal geometry from neighboring polymer chains
is negligible.

\bibitem{revPBE}
Y. Zhang and W. Yang, Phys. Rev. Lett. \textbf{80}, 890 (1998).

\bibitem{ijqc}
D.C. Langreth et al.,
Int. J. Quantum Chem. \textbf{101}, 599 (2005).

\bibitem{dacapo}
\textsc{Dacapo} from \texttt{http://wiki.fysik.dtu.dk/dacapo}

\bibitem{vdBildt}
D. Vanderbilt, 
Phys. Rev. B \textbf{41}, 7892 (1990).

\bibitem{PECrystalwithPressure}
M.S. Miao et al., 
J. Chem. Phys \textbf{115}, 11317 (2001).

\bibitem{ExpPolyEthylene}
J.F. Twistleton, J.W. White, and P.A. Reynolds, 
Polymer \textbf{23}, 578 (1982).

\bibitem{Bill}
F.W. Billmeyer, J. Appl. Phys. \textbf{28}, 1114 (1957).

\bibitem{ChoyElasticPE}
C.L. Choy and W.P. Leung, J. Polym. Sci., Polym. Phys. Ed.
\textbf{23}, 1759 (1985).

\bibitem{DampPot}
See Refs. 5--11 in Ref.~\onlinecite{bonog}.

\end{thebibliography}
\end{document}